\documentclass{PoS}       

\usepackage{graphicx}
\usepackage{xcolor}
\usepackage{todonotes}

\ShortTitle{Results and status of the EUSO-TA detector}

\title{Results and status of the EUSO-TA detector}

\author{\speaker{Lech Wiktor Piotrowski} for the JEM-EUSO collaboration\footnote{for collaboration list see PoS(ICRC2019)1177}\\
	RIKEN, Wako, Japan\\
	E-mail: \email{publ@lwp.email}}

\abstract{
 
EUSO-TA is a ground-based telescope, located at the Black Rock Mesa site of Telescope Array (TA), Utah, USA. The main aim of the instrument is observation of Ultra High Energy Cosmic Rays through detection of ultraviolet light generated by cosmic-ray showers. EUSO-TA consists of two, 1 m$^2$ Fresnel lenses with a field of view of about $11^{\circ} \times 11^{\circ}$. Light is focused on the Photo Detector Module composed of 36 Hamamatsu multi-anode photomultipliers, for a total of 2304 channels. During operations, the telescope is housed in a shed located in front of the TA Fluorescence Detector. We present the results from EUSO-TA observational campaigns performed in years 2015 and 2016, including detected cosmic rays, meteors and laser shots. Also shown are the details of current instrument setup and upgrade of the detector to phase II. 
}

\FullConference{36th International Cosmic Ray Conference -ICRC2019-\\                                                                                                                                              
	July 24th - August 1st, 2019\\                                                                                                                                                                     
	Madison, WI, U.S.A.}

\begin{document}

\maketitle

\section{Introduction}

JEM-EUSO is a space-borne mission dedicated to detecting cosmic rays of the highest energies \cite{bib:JEMEUSO}. It will observe the ultra-violet light from Extended Air Showers (EAS) generated by cosmic rays in the atmosphere with a Fresnel lens based optics and super-fast single photon counting camera. Unlike the existing, on-ground experiments, JEM-EUSO due to its distance from the surface of the Earth, would observe a much larger volume of the atmosphere, significantly increasing the number of collected events and, hopefully, localising their sources in the Universe.

\begin{figure}[hbt]
	\begin{center}
		\includegraphics[width=0.7\textwidth]{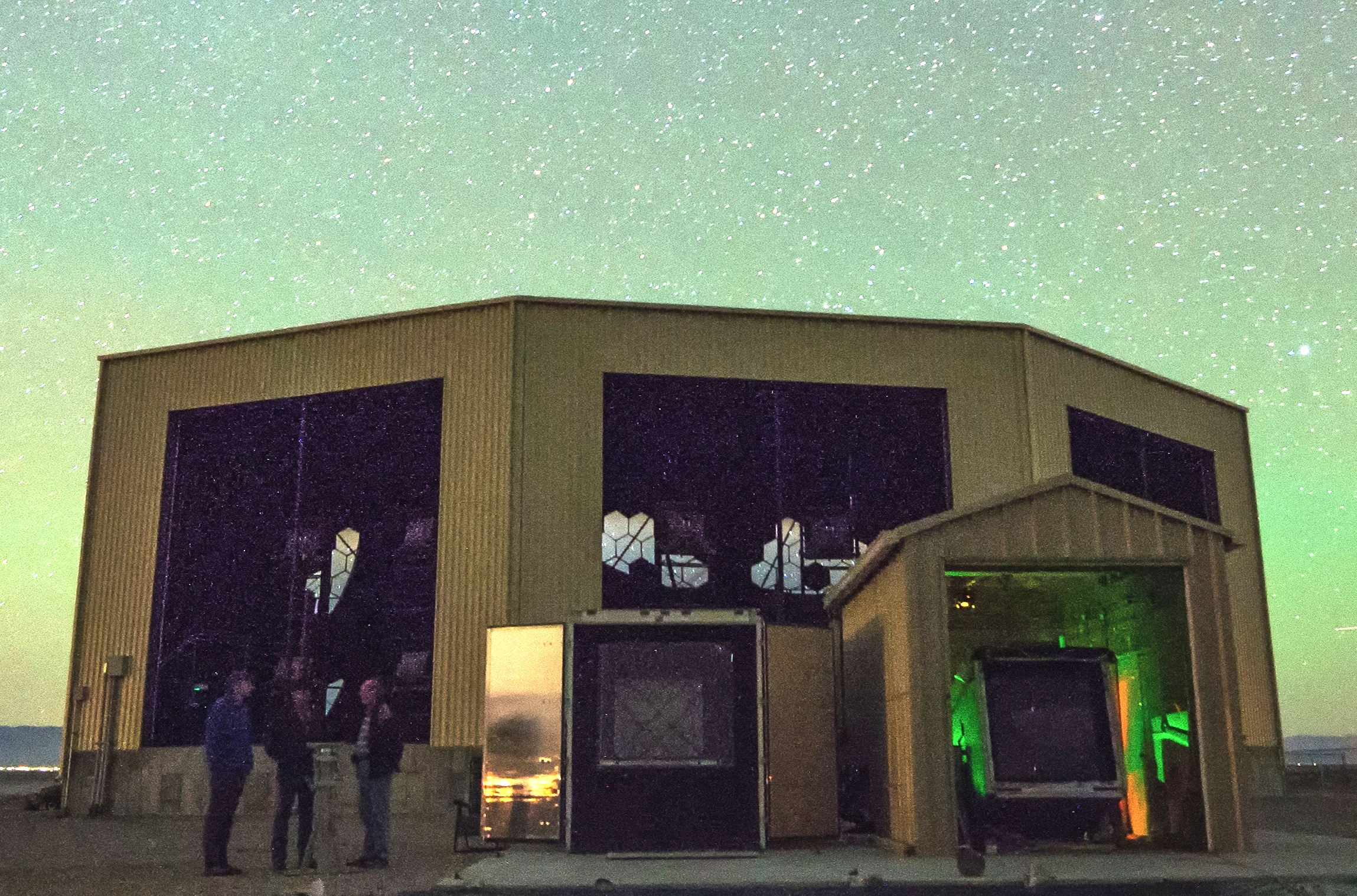}
		\caption{EUSO-TA (front right), EUSO-SPB \cite{bib:spb} (front middle) and Telescope Array Fluorescence Detector (TAFD, back) (photography by M. Mustafa)}
		\label{fig:TAFD}
	\end{center}
\end{figure}

\begin{figure}[hbt]
	\begin{center}
		\includegraphics[width=0.5\textwidth]{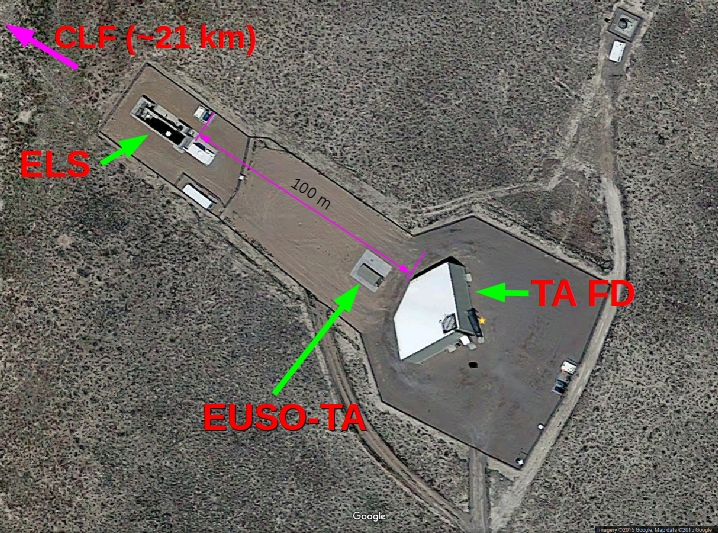}
		\caption{The position of EUSO-TA on the TAFD Site in Black Rock Mesa, Utah, USA. The EUSO-TA dome is placed just in front of the TAFD. The photograph was taken from Google Maps.}
		\label{fig:location}
	\end{center}
\end{figure}

EUSO-TA is a fully functional ground telescope (fig. \ref{fig:TAFD}) in EUSO family. It is located at Black Rock Mesa, Utah, at the site of one of the fluorescence light detectors of the Telescope Array (TA) experiment \cite{bib:TA}. From there it observes, simultaneously with TA, artificial light and cosmic ray events, allowing for tests of the technology, calibration of the detector and reduction of the systematic uncertainties of the measurements. The location and pointing allows for observation of TA's Central Laser Facility (CLF) and Electron Light Source (ELS) (fig. \ref{fig:location}). 

\section{EUSO-TA instrument}
\label{sec:telescope}

\begin{figure}[bt]
	\begin{center}
		\includegraphics[width=0.342\textwidth]{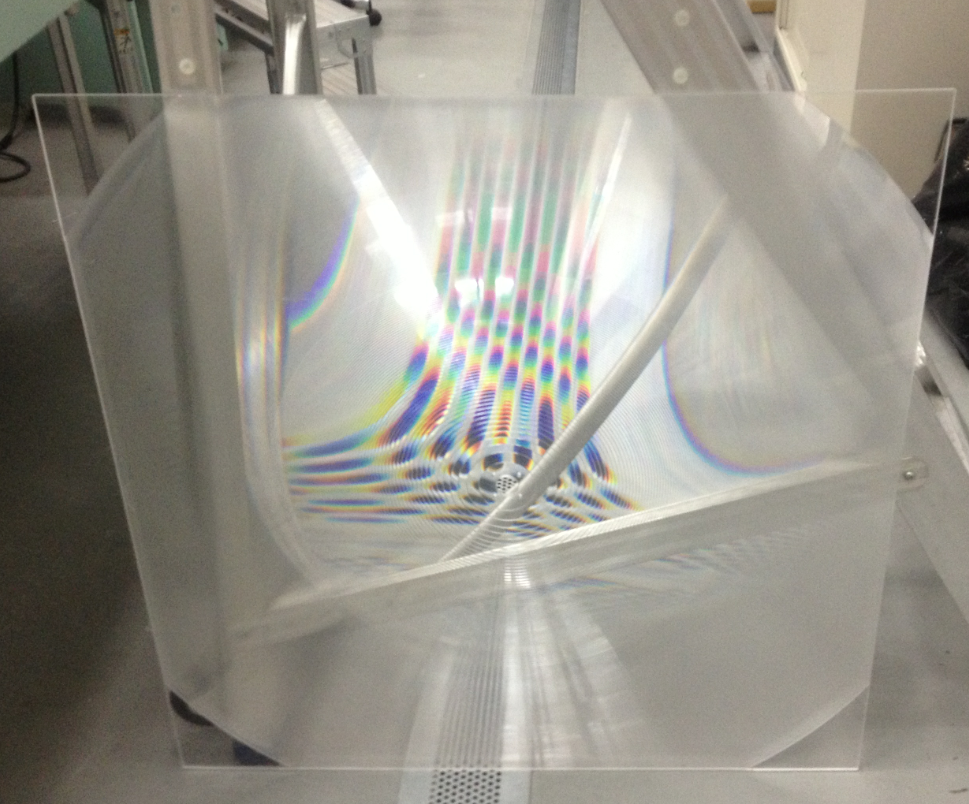}
		\includegraphics[width=0.3\textwidth]{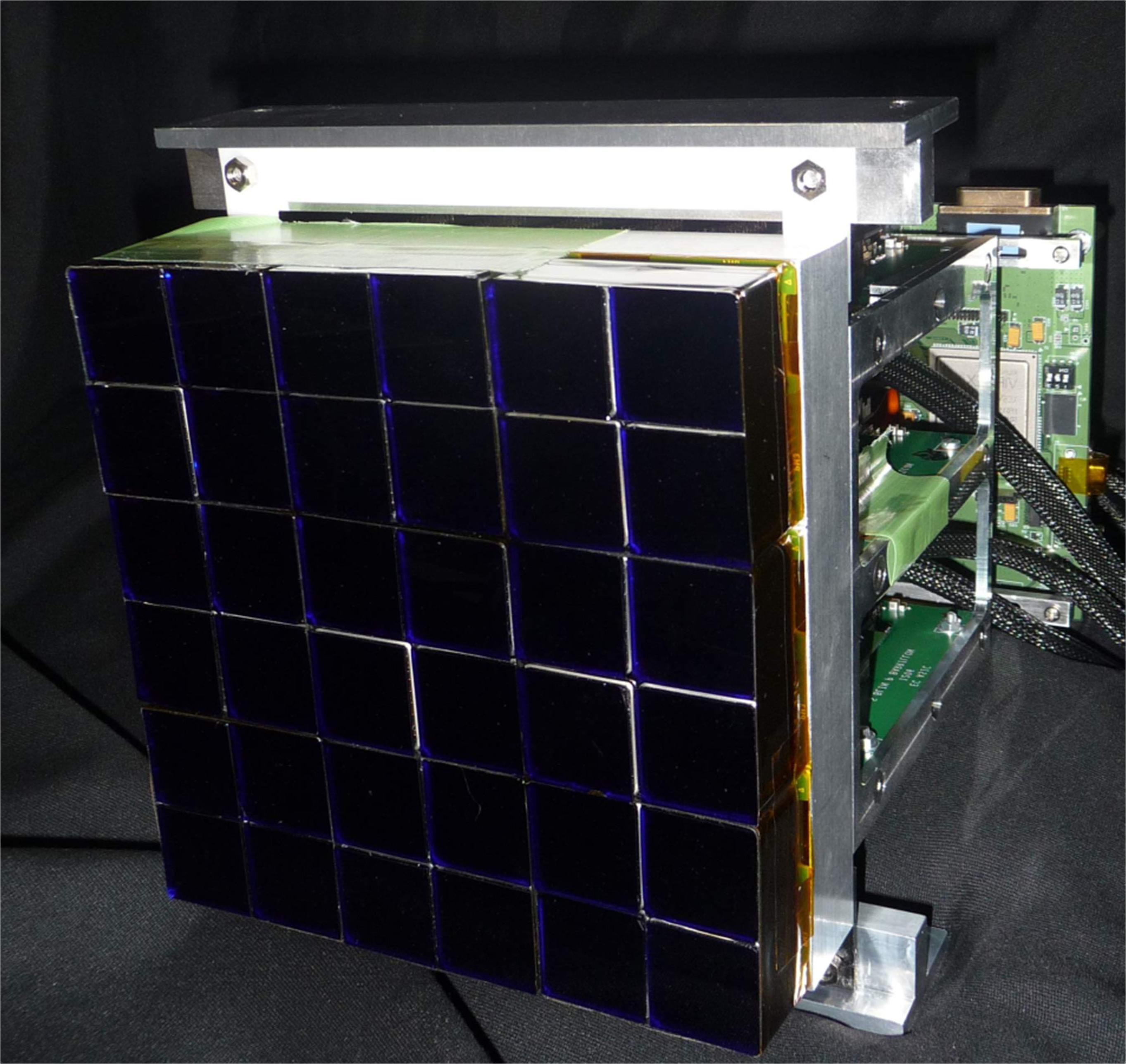}
		\caption{Front lens of EUSO-TA (left) and the PDM array and the front-end ASIC boards mounted on the PDM frame (right).}
		\label{fig:PMT_DP}
	\end{center}
\end{figure}

The 1 m$^2$ squared EUSO-TA Fresnel lenses are fabricated from UV transmitting polymethyl-methacrylate (PMMA). The lenses focus light on the 17 cm $\times$ 17 cm PDM, composed of 36 Multi-Anode Photomultiplier Tubes (MAPMTs) \cite{bib:mapmt_prieto} each containing 64 anodes, for a total of 2304 pixels (fig. \ref{fig:PMT_DP}). Four MAPMTs form an EC-Unit, each with a dedicated Cockroft-Walton based High Voltage Power Supply. The working voltage of the phototubes is 1000 V, and is automatically reduced in case of too bright light on an EC-Unit in single microseconds.

Each MAPMT is read out by one SPACIROC1 ASIC, which are distributed onto 6 EC-ASIC boards \cite{bib:ASIC}. The time resolution is $\sim30$ ns. This implies a saturation at about 28 counts on the frame -- the Gate Time Unit (GTU) -- of $2.3\ \mu s$. The GTU is followed by 200 ns of dead time.

The digitized counts from 6 EC-ASIC boards are read into a ring buffer of the PDM board. The buffer contains 128 GTUs and is read out on request, which can be made with an external or internal trigger. The data are then transferred to a Cluster Control Board (CCB), which contains a 2nd-level trigger \cite{bib:CCB} and CPU boards. Together with Clock board (CLKB), GPS, house keeping board and low voltage power supply \cite{bib:instrumental} they form a Data Processing (DP) unit. The exchange of information is made with encapsulated packets. The amount of information contained in a packet increases with the level of processing, the final packet stored by the CPU containing counts from ASICs, additional information from the PDM board, CCB, CLKB, GPS, and so on \cite{bib:euso-ta-data}.

\section{EUSO-TA campaigns}

The EUSO-TA lenses and mechanical structure was installed in Telescope Array Fluorescence Detector site in Black Rock Mesa, Utah, USA in March 2013. EUSO-TA PDM and readout system has been installed in February/March 2015. Since then, there were five observational campaigns of EUSO-TA, four in 2015 and one in 2016. More than 136 hours of observation was performed using TAFD external trigger, thus could contain UHCER events.

\section{Results}

\subsection{Stars observations}

\begin{figure}[b]
	\begin{center}
		\includegraphics[width=0.4\textwidth]{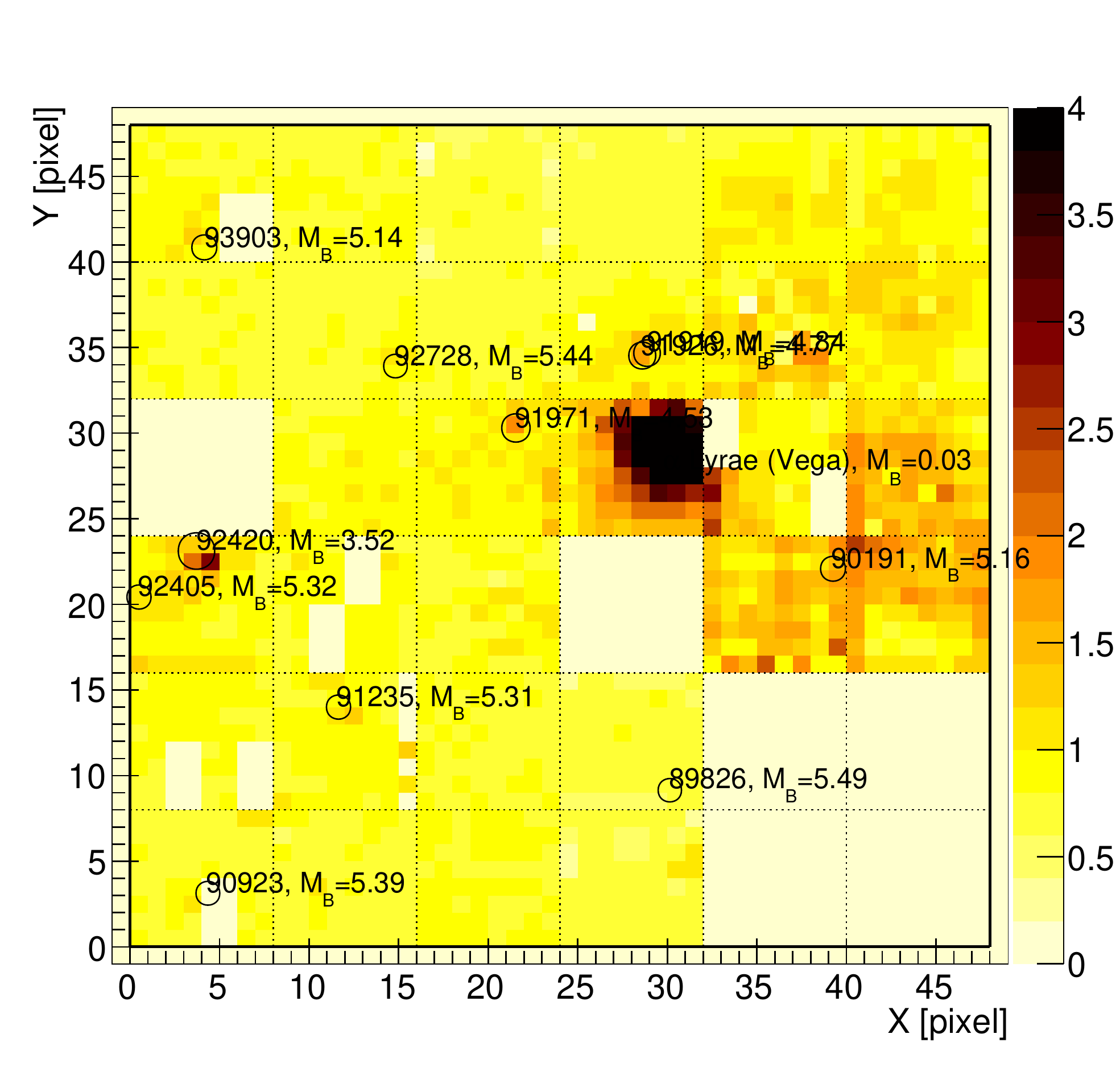}
		\caption{Sum of 1280 frames acquired with EUSO-TA with Hipparcos catalogue brightest stars' positions superimposed. The colour scale denotes the brightness of each pixel in arbitrary units after flat fielding.}
		\label{fig:stars}
		
	\end{center}
\end{figure}

EUSO-TA can observe stars up to magnitude in Johnson's blue filter $M_B\simeq6.5$ on sums of 1280 frames (about 3.2 ms observation time). This limiting magnitude is very approximate, since it depends largely on atmospheric conditions and stars spectra. An example of 1280 stacked frames with a few stars clearly visible is shown in fig. \ref{fig:stars}. The Hipparcos catalogue \cite{bib:hipparcos} is superimposed on the image for the 4 brightest stars, allowing us to recognize Algol and stars with $M_B\ge5.5$.

The stars can be used as point sources to analyze the point spread function (PSF) of our detector. Initial analysis performed with a fit of gaussian profile gives a PSF with average FWHM of 2.6 pixels \cite{bib:eta_psf}. It is well within the requirements for UHECR showers observations. Stars were also used to determine the field of view of the instrument, which is $10.6^{\circ} \pm 0.3^{\circ}$.

\subsection{Other ``slow'' signals}

\begin{figure}[h!]
	\begin{center}
		\includegraphics[width=0.4\textwidth]{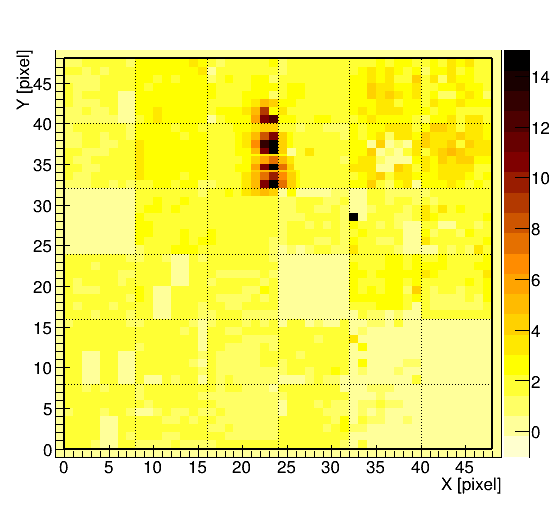}
		\caption{A meteor track detected by EUSO-TA. The picture shows an overlap of four averages of 1280 frames. The color scale denotes the uncalibrated detector counts.}
		\label{fig:meteor}
		
	\end{center}
\end{figure}

While EUSO-TA is designed for observations of millisecond-scale events, it can observe phenomena happening on much longer timescales. The most numerous are flashes from airplanes and sunlight reflected by satellites. EUSO-TA has also observed meteors, as can be seen in fig. \ref{fig:meteor}. Such observations are unique due to their time resolution. It could be important in a case of detecting a phenomenon similar in appearance -- strangelets \cite{bib:nuclearites}.

\subsection{Lasers observations}
\label{sec:laser}

To study the EUSO-TA response to a known light source we have used the light coming from the TA's CLF, distant from EUSO-TA by about 21~km. The CLF shoots vertically a laser of 355 nm wavelength in front of the detectors\cite{bib:clf}.

\begin{figure}[h!]
	\begin{center}
		\includegraphics[width=0.325\textwidth]{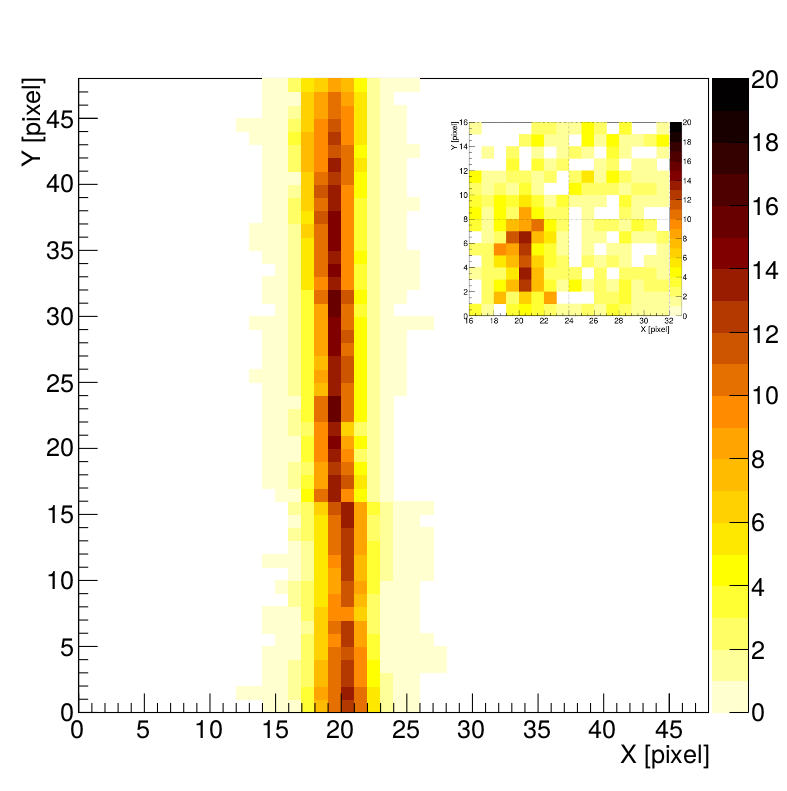}
		\includegraphics[width=0.49\textwidth]{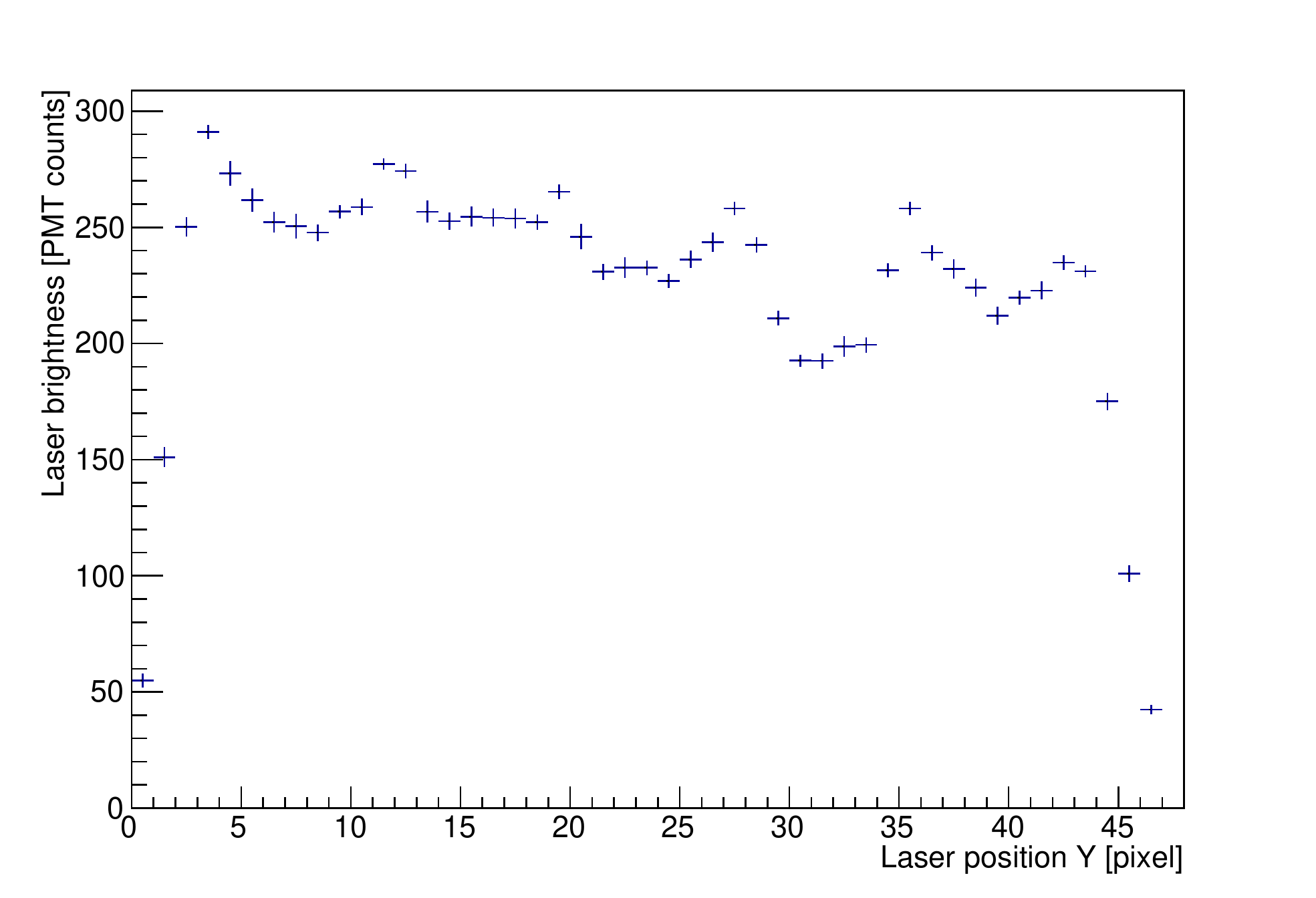}
		\caption{Left: an average of 259 tracks of CLF laser. The subfigure shows a zoomed part of a single frame containing the laser spot. The colour scale denotes the uncalibrated counts. Right: a profile histogram of 259 CLF tracks. Each point value shows an average of summed counts of all tracks at a corresponding vertical position of shots. The brightness is lower for spots falling into space between MAPMTs or out of the PDM.}
		\label{fig:lasers}
	\end{center}
\end{figure}

During standard observation nights the CLF is shot every half an hour for 30 s with 10 Hz shooting frequency. The scattered light of the $\sim 3$ mJ beam is clearly visible traversing through the EUSO-TA field of view on 6 to 8 frames, depending on the shot and acquisition time synchronization. The spot length is 6-8 pixels depending on the position on the frame (fig. \ref{fig:lasers}, left), which is consistent with expectations. The registered light intensity is dependent both on the atmospheric conditions and the fine features of the detector's focal surface.

In addition to the CLF we have also performed measurements of a Global Light System (GLS) laser - a mobile UV laser of Colorado School of Mines. The laser can be shot with energies in the range of about 1--86 mJ, with pointing adjustable in two dimensions. The mechanics featured automatic changing of the pointing, allowing for easy ``swipes'' through the field of view and emulating the orbital condition where the UHECR increases its distance to the detector with time.

\begin{figure}[bt]
	\begin{center}
		\includegraphics[width=0.49\textwidth]{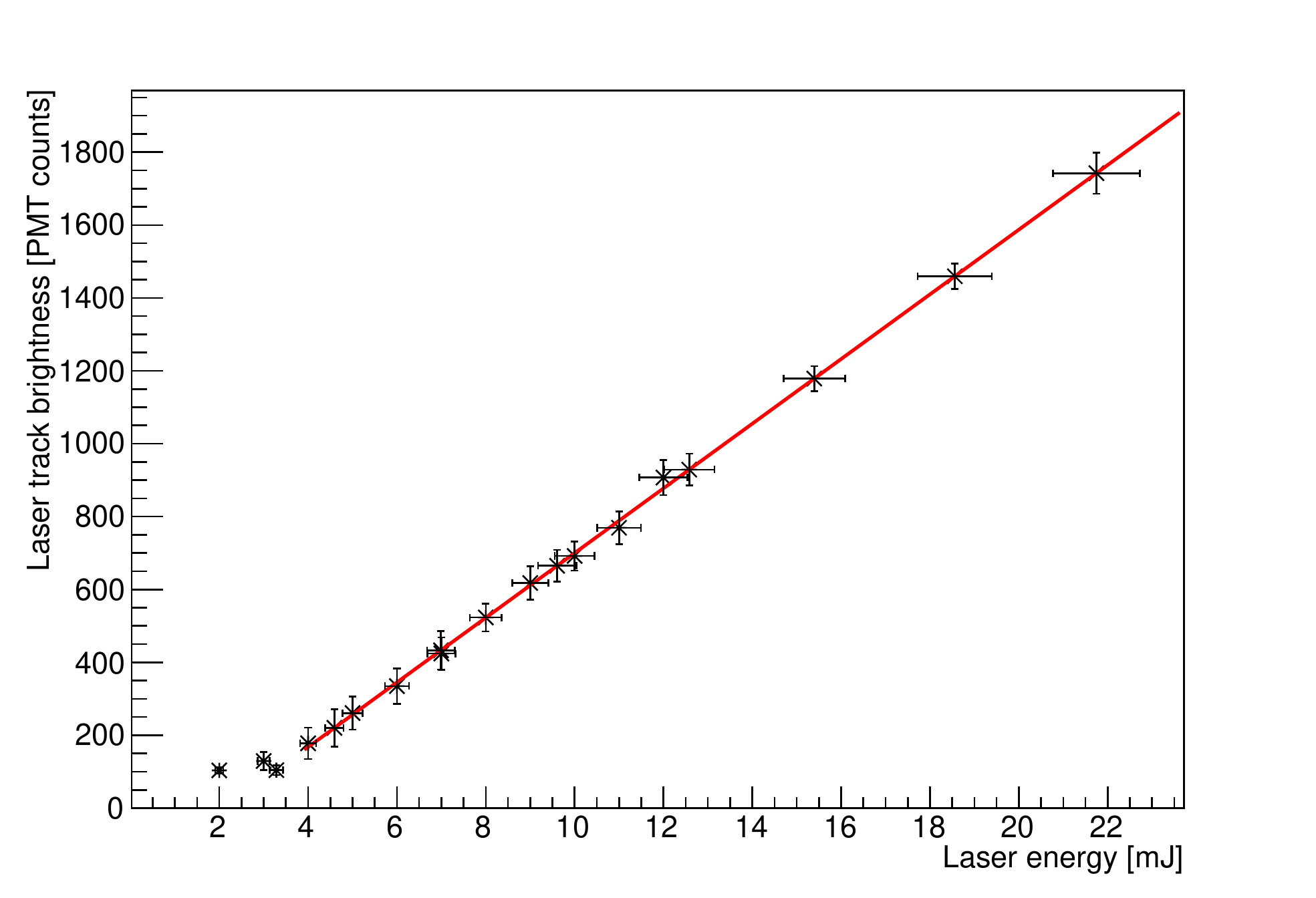}
		\caption{Reconstructed brightness of the GLS laser track vs its energy for the distance of 33 km from the detector. The plot is joined from 2 shooting sessions, altogether encompassing energy range of 2--22 mJ. Each point was calculated from a few dozen shots. Line fit shows good linearity of the detector response in the range 4--22 mJ, while for lower energies only the brightest tracks were reconstructed, enforcing a low brightness cut on the data points. }
		\label{fig:laser_linearity}
	\end{center}
\end{figure}

Fig. \ref{fig:laser_linearity} shows reconstructed brightness of the laser track in the detector vs laser shot energy, for the distance of 33 km. In the tested energy range of 4--22 mJ, which is outside the saturation region, the dependency is linear, showing that the detector behaves as expected. For the lowest energies of 2--3 mJ, only the brightest laser events were reconstructed, resulting in a brightness cut on the data.

\subsection{UHECR}

\begin{figure}[h!]
	\begin{center}
		\includegraphics[width=0.45\textwidth]{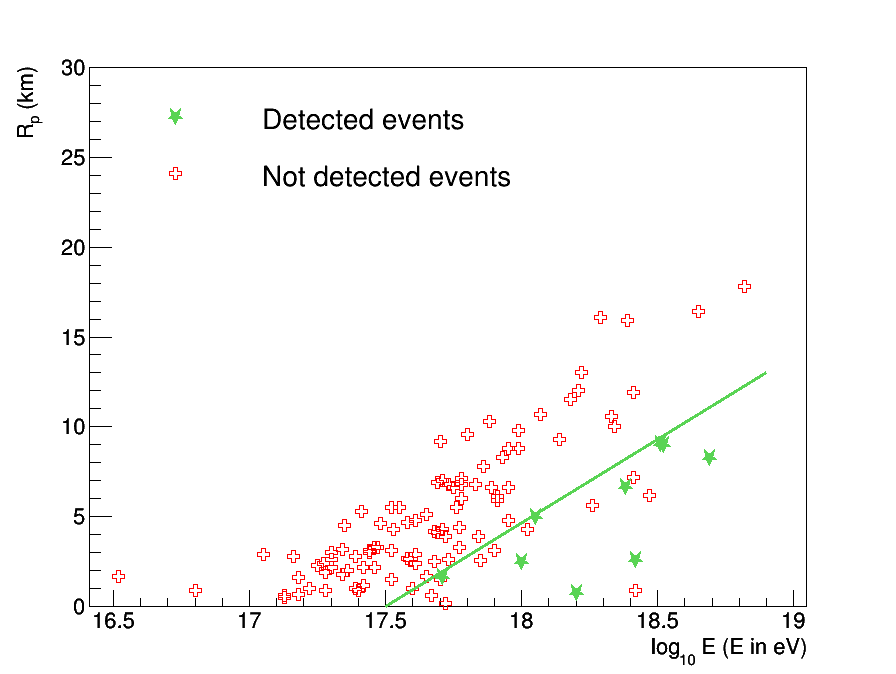}
		\caption{All UHECR registered by TAFD in the EUSO-TA field of view during its operation. Markers' shape and colour denotes status in EUSO-TA: not detected, detected and detected in simulation, with a distance-energy sensitivity border emerging from the data.}
		\label{fig:uhecr_statistics}
	\end{center}
\end{figure}

\begin{figure}[t]
	\begin{center}
		\begin{tabular}{cc}
			\small Data & \small Simulation \\
			\includegraphics[width=0.3\textwidth]{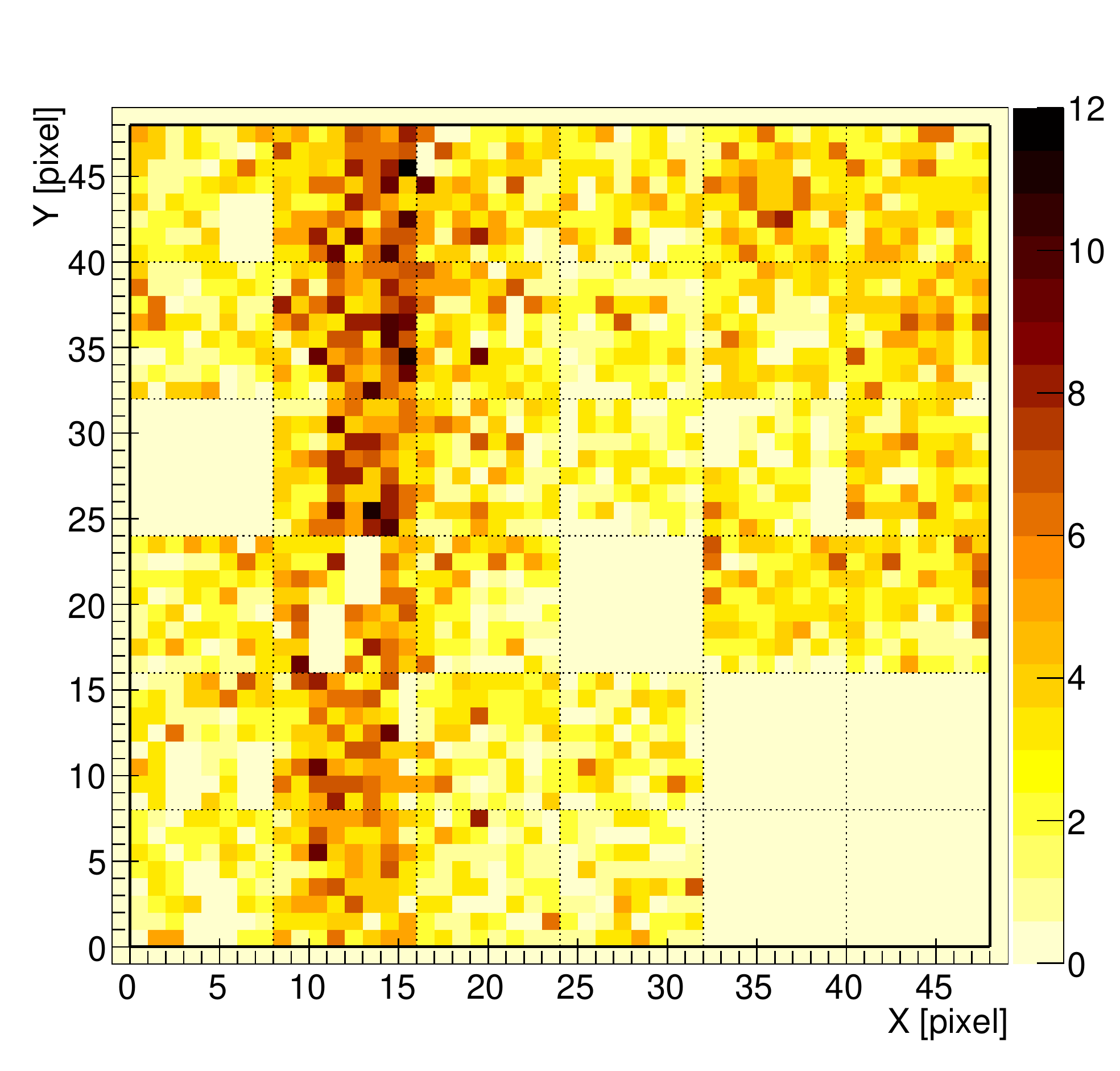} & 
			\includegraphics[width=0.3\textwidth]{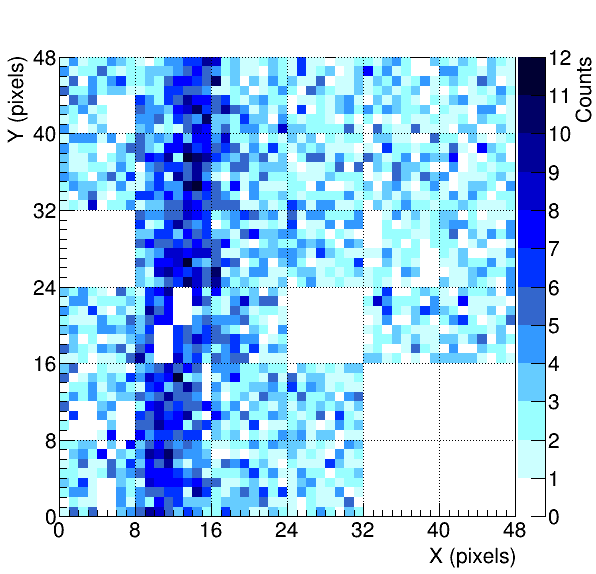}	\\		
		\end{tabular}
		\caption{Example of an UHECR observed by EUSO-TA. Left shows the real data in photoelectron counts, right the simulation. The shower had an energy of $10^{18.36}$ eV and was 2.6 km from the telescope (data from TA measurements).}
		\label{fig:uhecr_event}
	\end{center}
\end{figure}

To date, nine UHECR events have been identified in $\sim 130$ hours of UHECR-dedicated observations (fig. \ref{fig:uhecr_statistics}). Their distances from the detector vary between 0.8 and 9 km, while the energy between $10^{17.7}-10^{18.8}$ eV, according to TA measurements. An example event is shown in fig. \ref{fig:uhecr_event}.

These first events registered with EUSO technology allowed us to vastly improve our reconstruction and simulation algorithms, the work still being in progress. However, EAS parameters had to be derived from TAFD which, thanks to larger FoV and higher time-resolution, could see the shower movement. The closer look at the parameters of UHECR that happened in EUSO-TA field of view during its operation starts to reveal the detection capabilities of the detector \cite{bib:eusota_events}.

\section{EUSO-TA future plans}

EUSO-TA is the best available way to test technology for existing and future EUSO family experiments, as it allows for stable field observations for extended time periods. However, the observation time will be significantly increased with automatisation of the telescope, which will allow remote operations. In addition, we are refurbishing our focal surface, upgrading ASIC to SPACIROC3 with lower dead time and higher sensitivity, as well as changing the FPGA board to a Xilinx Zynq based one, which enables faster and more sophisticated data processing, including new autonomous triggers for cosmic rays \cite{bib:trig_mat} and slow events such as meteors, strangelets or lightnings, which are mostly discarded by the TAFD. This will expand the scientific goals of EUSO-TA.

\section{Summary}

EUSO-TA employs a new technology of observing cosmic rays, with use of Fresnel lenses and multi-anode photomultipliers. The detector has registered nine UHECRs during its five observational campaigns, proving that this approach works well for the main goal. Additionally, a number of ``slow'' events such as stars, meteors and planes have been observed allowing for an extension of scientific objectives.

The main goal of the detector, however, was the test of hardware capabilities. It proved invaluable in the modifications applied for EUSO-SPB and Mini-EUSO \cite{bib:mini-euso} detectors. In near future we plan an upgrade and automatisation of the telescope to increase the sensitivity and duty cycle.

\begin{acknowledgments}

This work was partially supported by Basic Science Interdisciplinary Research Projects of RIKEN and JSPS KAKENHI Grant (22340063, 23340081, and 24244042), by the Italian Ministry of Foreign Affairs and International Cooperation, by the Italian Space Agency through the ASI INFN agreement n. 2017-8-H.0, by NASA award 11-APRA-0058 in the USA, by the French space agency CNES, by the Deutsches Zentrum fur Luft-und Raumfahrt, the Helmholtz Alliance for Astroparticle Physics funded by the Initiative and Networking Fund of the Helmholtz Association (Germany), by Slovak Academy of Sciences MVTS JEM-EUSO, by National Science Centre in Poland grant (2015/19/N/ST9/03708), by Mexican funding agencies PAPIIT-UNAM, CONACyT and the Mexican Space Agency (AEM), as well as VEGA grant agency project 2/0132/17, and by State Space Corporation ROSCOSMOS and Russian Foundation for Basic Research (grant 16-29-13065). We are grateful to the Telescope Array collaboration for all their help.

\end{acknowledgments}

{}

\end{document}